# Water affinity to epitaxial graphene: the impact of layer thickness

*Cristina E. Giusca[1]\*, Vishal Panchal[1], Martin Munz[1], Virginia D. Wheeler[2], Luke O. Nyakiti[3], Rachael L. Myers-Ward[2], D. Kurt Gaskill[2], Olga Kazakova[1]*

[1]National Physical Laboratory, Hampton Road, Teddington, TW11 0LW, United Kingdom
[2]U.S. Naval Research Laboratory, Washington, DC 20375, United States of America
[3]Texas A&M University, Galveston, TX 77553, United States of America

**ABSTRACT**: The sensitivity to water of one-, two- and three-layer epitaxial graphene (1, 2 and 3LG) is examined in this study. We unambiguously show that graphene's response to water, as measured by changes in work function and carrier density, is dependent on its thickness, with 1LG being the most sensitive to water adsorption and environmental concentration changes. This is furthermore substantiated by surface adhesion measurements, which bring evidence that 1LG is less hydrophobic than 2LG. Yet, surprisingly, we find that other contaminants commonly present in ambient air have a greater impact on graphene response than water vapour alone. This study indicates that graphene sensor design and calibration to minimize or discriminate the effect of the ambient in which it is intended to operate are necessary to insure the desired sensitivity and reliability of sensors. The present work will aid in developing models for realistic graphene sensors and establishing protocols for molecular sensor design and development.

---

\* To whom correspondence should be addressed. E-mail: cristina.giusca@npl.co.uk

Although graphene has been shown to exhibit great potential for sensing applications [1, 2, 3], no attention is paid to how the thickness of graphene affects the sensor response in terms of sensitivity, selectivity, response time or reproducibility. The different electronic structure of one-, two- and thicker graphene layers [4] suggests that the sensor response can be affected by domain thickness and we demonstrate this through subjecting graphene samples with well-defined number of layers to varying humidity levels. Since water is the most abundant dipolar adsorbate under ambient[†] conditions, significant effort has been dedicated to both theoretical [5, 6, 7] and experimental [8, 9, 10] investigations of water on graphitic surfaces in an attempt to elucidate the water-graphene interaction from an electronic structure perspective. While some density functional theory (DFT) calculations predicted the formation of a small energy gap of the order of 20-30 meV when graphene is fully covered with water molecules [16], other theoretical studies show that adsorbed water has very little effect on the electronic structure of graphene [7, 17]. The unpredictable influence that environmental humidity exposure have on graphene devices processed and operated in ambient has also been highlighted by several previous experimental studies [11, 12, 13, 14, 15]. Nevertheless, there is currently a lack of studies concerned with quantitative aspects and reliable measures of changes in the electronic properties of graphene due to environmental humidity changes. Furthermore, studies of graphene exposed to various target gases (not commonly found in the ambient in the concentrations of interest) and ambient environment have been carried out on mechanically exfoliated graphene [9, 10, 19], with very little work done on epitaxial graphene, which holds a practical route for many applications (e.g. optoelectronics, Quantum Hall effect metrology, high speed electronics), given its compatibility with wafer-scale processing techniques. Moreover, the intrinsic doping level information obtained from these transport measurements is generalized over the entire device and is not correlated with the exact morphology of graphene in terms of domain thickness and the presence of local adsorbates that can influence the electronic properties of graphene.

---

[†] Here, ambient refers to the typical work place or laboratory environment, that is, the air has not been specially processed or cleaned of impurities. 4.79For this study, ambient is the untreated air in the laboratory where the samples were investigated.

In the current work, we employ scanning Kelvin probe microscopy (SKPM) to investigate the effect that water has on the electronic properties of epitaxial graphene, directly correlated with local structural information and electrical transport measurements, in order to provide a systematic evaluation of the impact of ambient exposure on epitaxial graphene, in accordance with its domain thickness. SKPM provides an electrical map of the surface that gives essential information on graphene thickness, layer-dependent distribution of charge, electrical potential and work function [20, 21, 22]. We study the influence of relative humidity variation (RH=10-70%) on the surface potential of one-, two- and tri-layer epitaxial graphene (1LG, 2LG and 3LG, respectively) on SiC, as well as the effect that the change in environment, from ambient to vacuum and to high humidity levels, has on the electronic properties of the various graphene domains. Additionally, the water-graphene interaction is examined in terms of water adhesion to 1LG and 2LG, with a view to correlating the surface potential with the degree of wettability of graphene domains of different thicknesses.

We unambiguously demonstrate that graphene domains of different thicknesses react differently to the change in environment and that the sensitivity to water, as measured by changes in work function and carrier density, increases with decreasing domain thickness, with 1LG being the most sensitive to water adsorption and change in environment. Understanding the role of 1, 2 and 3LG and their response to various environments will impact the design and development of molecular sensors and will help to correctly model the response of a realistic sensor.

Epitaxial graphene films have been investigated by Raman spectroscopy, SKPM and adhesion measurements, with Hall bar devices for transport measurements fabricated using standard electron beam lithography. SKPM experiments on graphene films and transport measurements of graphene devices have been performed in ambient (for "ambient" definition see above), vacuum, nitrogen and increasing humidity levels, as described further below. Details of sample growth, device fabrication and transport measurements, SKPM, adhesion measurements and Raman spectroscopy measurements are given in Methods.

**Raman spectroscopy:** In order to evaluate the quality and number of constituent layers, Raman spectroscopy mapping has been carried out and provides evidence that the sample consists mainly of 1LG with continuous strips of 2LG formed along SiC terrace edges, as exemplified in Figure 1a and 1b, where G-band and 2D-band intensity maps of a (15×15) µm² area are presented.

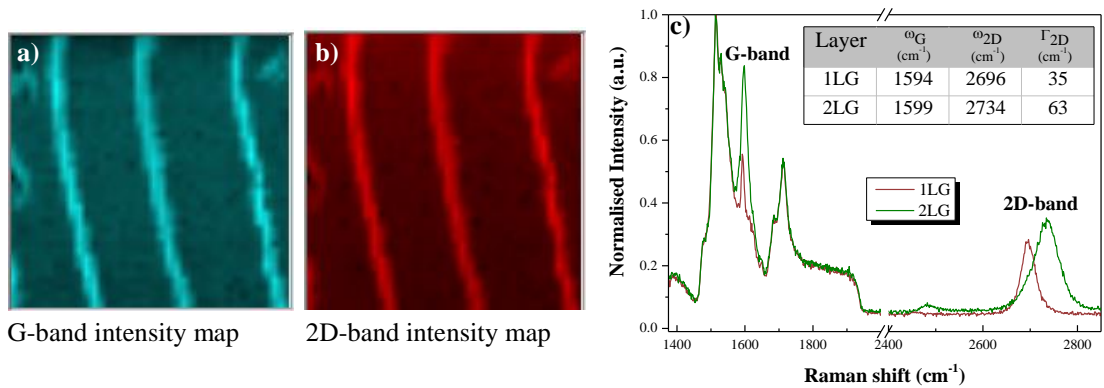

**Figure 1: G-band (a) and 2D-band (b) Raman relative intensity maps of (15x15) µm² area graphene obtained using 532-nm excitation wavelength. (c) Representative individual Raman spectra for 1LG and 2LG. In the G-band range the spectra are dominated by the SiC modes of the supporting substrate. Inset table shows the G and 2D peak position, $\omega_G$ and $\omega_{2D}$, for 1LG and 2LG, respectively, and the full-width-at-half-maximum $\Gamma$ for the 2D band. SiC contribution was not subtracted from the individual spectra to highlight that changes in peak positions and widths are only associated with graphene and not with SiC modes.**

As illustrated in Figure 1c, representative individual Raman spectra for 1LG and 2LG display both the G- and 2D-modes typical to graphene, as well as the second order features of SiC in the range 1450 - 1750 cm$^{-1}$. The G-band, associated with the doubly degenerate phonon mode vibrations at the center of the Brillouin zone, is located at ~1594 cm$^{-1}$ for 1LG, but it becomes more intense, twice wider and shifts to higher wavenumbers for 2LG. A similar trend is observed for the 2D-band of 2LG that displays a blue shift compared to that of 1LG, while becoming more asymmetric and almost doubling its width. Moreover, the shape of the 2D-band for 1LG can be perfectly fitted using a single Lorentzian peak, whereas four distinct components are required for that of 2LG, as expected with the evolution of the electronic bands on the transition from one to two graphene layers [23].

**Scanning Kelvin probe microscopy:** SKPM experiments were conducted in an environmental chamber, where the atmosphere was controlled using the sequence: ambient (as defined above), vacuum, nitrogen and increasing humidity levels in nitrogen, on the same region of the sample and using the same scanning tip to ensure consistency of measurements. Topography was recorded simultaneously with surface potential images in all cases. As no modification of topography had been observed throughout the entire sequence of measurements, only the image recorded in ambient is shown for exemplification in Figure 2a. A sequence of representative surface potential images of epitaxial graphene characterised by in-situ SKPM under various environments (ambient 1, vacuum, nitrogen, humid conditions and ambient 2) is displayed in Figure 2 (b-l). The initial measurements were carried out in ambient (ambient 1), at room temperature (T = 21°C) with no prior surface conditioning of the graphene sample. Following ambient exposure, the surface potential was measured in vacuum after the sample was annealed at 150°C to remove water and atmospheric adsorbates and subsequently cooled down to room temperature. The following step involved exposing the sample to nitrogen gas (99.9995% purity) in order to bring the environmental chamber from vacuum to atmospheric pressure for surface potential measurements under humid conditions. The humidity exposure sequence was carried out in a stepwise manner, from 10% RH to 70% RH, using N$_2$ as a background gas. Finally, the sample was re-exposed to ambient (ambient 2) after the entire measurement cycle. Further details on exposure to various environments and measurement time are included in Methods.

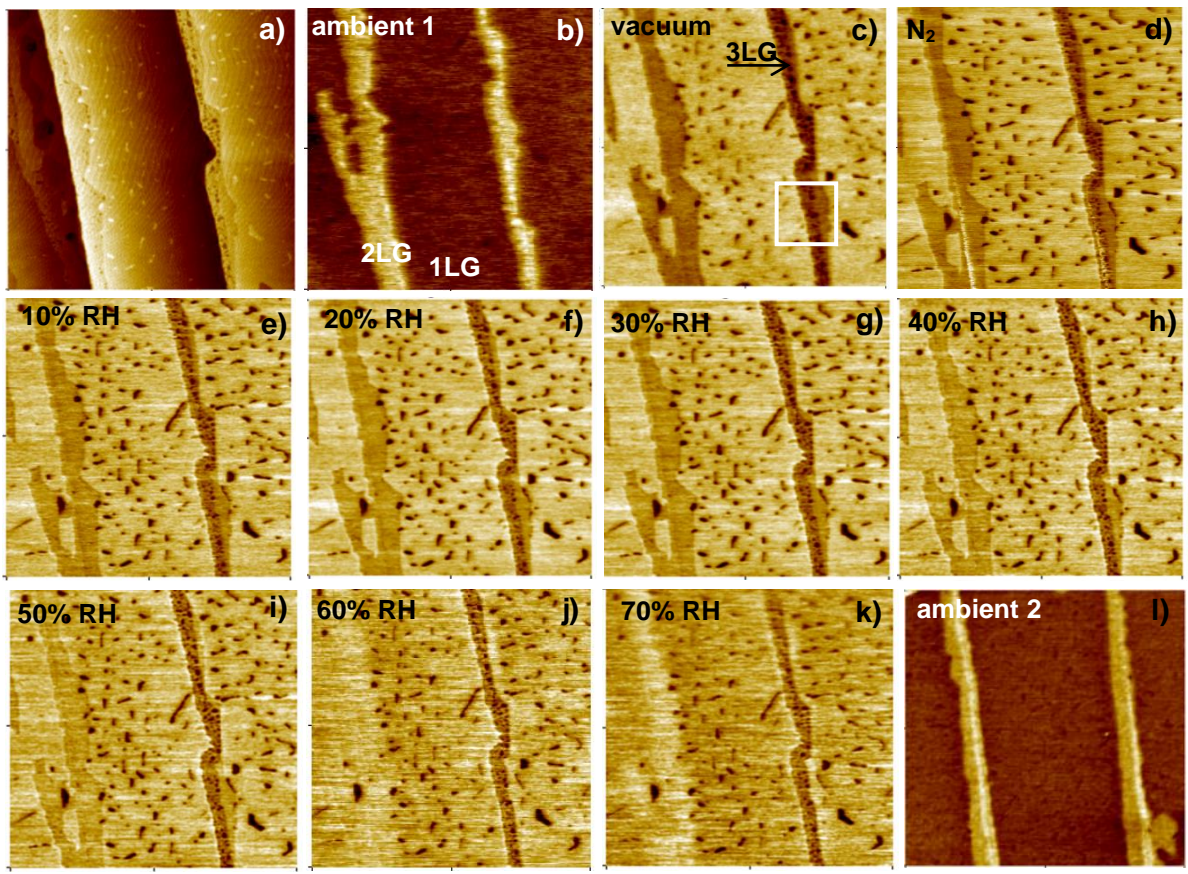

**Figure 2:** Topography (a) and sequence of surface potential images (b-k) collected on the same region of the graphene sample, while environmental conditions were changing from ambient to vacuum, nitrogen and to increasing levels of humidity in nitrogen. Image in (l), showing the surface potential after re-exposure to ambient, was taken on a different region of the sample due to a small offset induced by drift when venting the environmental chamber. The scan size is (10x10) µm² for all images. Surface potential images were acquired simultaneously with topography images, but only one representative topography image is shown in (a). The surface potential maps are plotted on different Z scales to enhance difference in contrast. See Table S1 for the absolute values of surface potential for 1, 2 and 3LG.

Topography image displayed in Figure 2a reveals a terraced surface morphology with parallel edges consistent with typical SiC morphology, also showing the scarce presence of buffer layer regions as small, bright filamentary structures on terraces. While topography does not allow for a clear identification of the number of graphene layers, the surface potential map acquired simultaneously in ambient conditions (Figure 2b), shows regions with two main distinct contrast levels: a bright one, given by two parallel stripes, associated with 2LG, superimposed on a dark contrast background of 1LG. This is furthermore highlighted by the corresponding histogram associated with the surface potential map (Figure 3a), displaying a bi-modal surface potential distribution, consistent with the number of layers revealed by Raman spectroscopy maps in Figure 1a. The histogram presents the surface potential distribution relative to 1LG, which was artificially assigned to 0 eV, in order to emphasise the surface potential difference between the layers. Peak deconvolution was carried out using Lorentzian shape components and layer thickness was designated in accordance with the domains contrast in the surface potential image, such that the most intense peak corresponds to the brightest contrast in the image.

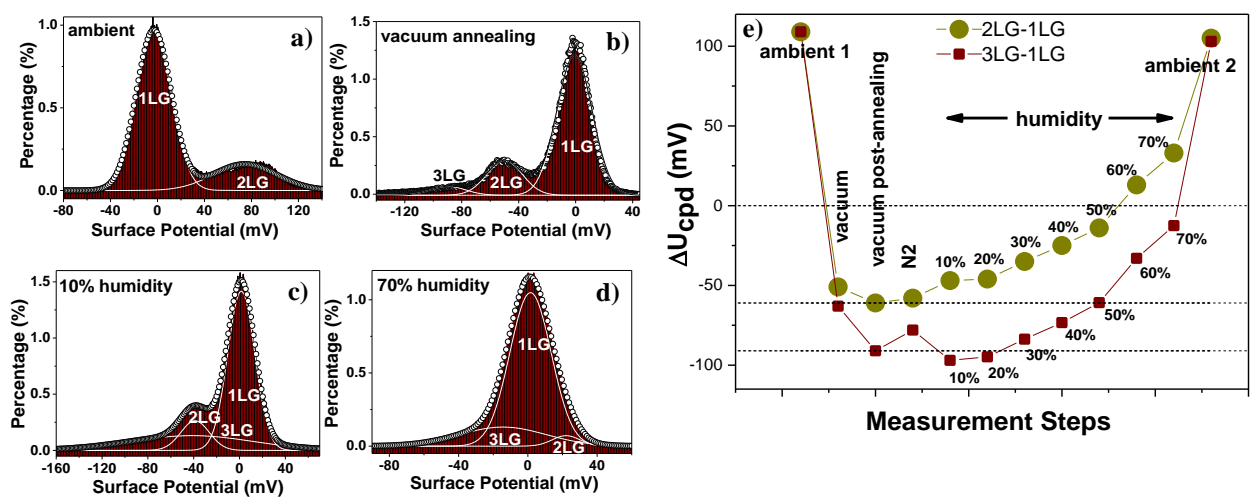

**Figure 3: Histograms associated with surface potential maps shown in Figure 2, illustrating relative contact potential difference ($U_{CPD}$) values between individual layers,**

**are shown for ambient (a), vacuum annealing (b), low humidity (c), and high humidity (d) conditions, respectively. Histograms correspond to the area enclosed by the white square in Figure 2c. Curve fitting and layer thickness designation are described in text. (e) Sequence of contact potential difference ($\Delta U_{CPD}$) variations for 2LG and 3LG with respect to 1LG. Environmental conditions were changing in the following order: ambient, vacuum, nitrogen, increasing levels of humidity, from 10% to 70% RH, followed by re-exposure to ambient. "Ambient 1" and "ambient 2" represent the values recorded at the initial and the final step of the measurement cycle, respectively. The value of ambient humidity was 40% RH.**

Based on the values derived for the contact potential difference in ambient and calibrated value for the tip work function, $\Phi_{tip}$ = 4.1 eV (see Methods) [23], work function values for 1LG and 2LG have been calculated using $e\Delta U_{CPD} = \Phi_{tip} - \Phi_{surface}$. The work function values obtained for ambient are 4.79 eV for 1LG and 4.68 eV for 2LG. Previous SKPM studies in ambient report a work function value of ~ 4.5 eV for graphene [24, 25, 26], that depends on the number of layers and can be tuned by ion irradiation-induced defects [25] or electric field effect within the range 4.5 – 4.8 eV for 1LG and 4.65 – 4.75 eV for 2LG [26]. Following ambient exposure, the chamber was pumped for ~16 hours to P = $1\times10^{-5}$ mbar and subsequently annealed at 150°C for 2 hours, under vacuum conditions, in order to remove impurities and air adsorbates, including water. This procedure always results in a reproducible surface, as suggested by the relatively constant surface potential values obtained for multiple measurement cycles [18], and is consistent with previous experiments reporting reduction of p-doping following vacuum annealing of graphene [12, 27]. Contrast inversion of the surface potential (with respect to surface potential observations in ambient conditions) is observed upon vacuum annealing (Figure 2c) and the related histogram (Figure 3b), where 1LG now displays brighter contrast than 2LG. Interestingly, 3LG becomes visible in the surface potential map under vacuum and shows an even darker contrast relative to 1LG than 2LG. Additionally, the buffer layer features (hardly distinguishable on the surface potential image under ambient conditions) have also become visible in vacuum. The work function value of 1LG obtained in vacuum, $(4.31 \pm 0.02)$ eV agrees well with the value of 4.34 eV measured by A. Tadich et al. on clean epitaxial graphene [28].

In the next set of experiments, after vacuum annealing the sample was first exposed to dry nitrogen at atmospheric pressure and then to varying humidity levels, ranging between 10% and 70% RH. Change in contrast is observed in the surface potential images for 2LG and 3LG relative to 1LG with the change of environment and RH, as seen in Figure 2 e-k, whereas associated topography images (not shown) remain unaffected.

The summary of these experiments is presented in sequential order in Figure 3e, illustrating the contact potential difference $\Delta U_{CPD}^{2-1}$ and $\Delta U_{CPD}^{3-1}$ measured for 2LG and 3LG relative to 1LG, respectively. $\Delta U_{CPD}^{2-1}$ and $\Delta U_{CPD}^{3-1}$ values for each environment (ambient, vacuum, nitrogen, RH in $N_2$) have been extracted from associated histograms of surface potential distribution, as shown in Figure 3a-d. The surface potential values for 1, 2 and 3LG for each environment are presented in Table S1 in the Supporting Information section. The plot in Figure 3e shows that the contact potential difference measured by SKPM between the 2LG and 1LG, $\Delta U_{CPD}^{2-1}$ changes sign, from ~109 mV in ambient to ~ -51 mV in vacuum, which decreases further to ~ -61 mV upon annealing in vacuum. This behaviour is consistent with gradual desorption of loosely bound species, such as water and other electron withdrawing adsorbed species (p-dopants), which appear to derive from the ambient, during pumping. However, pumping alone is not sufficient to remove all environmental dopants (or at least for the 16 hours used in this study). As shown by the further decrease in surface potential, heat treatment of the sample in vacuum (for 2 hours, at 150°C) seems to be more effective in removing environmental adsorbates than pumping alone. Our observations are in agreement with previous reports showing that the characteristics of four-terminal graphene devices improve upon pumping and heat treatment, resulting in a smaller hysteresis in the electric field effect behaviour, charge neutrality point closer to zero back-gate voltage and larger resistance at the charge neutrality point [12], [40].

Introducing $N_2$ into the chamber after vacuum annealing the sample and prior to humidity exposure, shows a slight increase of $\Delta U_{CPD}^{2-1}$ to ~ -58 mV. $N_2$ is an inert gas at room temperature, and it should not produce any change in surface potential, however contaminants can be present on the inner walls of the non-metallic tube carrying $N_2$ into the chamber, even though flushing the line with $N_2$ gas had been carried out several times. Following this step, the absolute value of $\Delta U_{CPD}^{2-1}$ gradually increases with increasing RH, passes through 0 at ~ 55% RH and reaches positive values for 60% and 70% RH levels. The work function values obtained for 70% RH are 4.59 eV for 1LG and 4.55 eV for 2LG. As illustrated in Figures 2j and 2k, 2LG shows inverted contrast at 60% and 70% RH, although not reaching the level observed initially under ambient conditions.

A similar trend is observed for $\Delta U_{CPD}^{3-1}$ measured for 3LG with respect to 1LG, as illustrated in Figure 3e, although with more depressed values that do not cross the zero-level to invert the contrast for 3LG.

It is important to point out that the surface potential (and consequently the work function) values reproducibly return to the initial level when the sample is re-exposed to ambient following the measurement cycle in the various environments. The work function variation with the change in environment for 1, 2 and 3LG is displayed in Figure 4 and the respective work functions are summarised in Table S1. The work functions have been obtained based on the absolute values of surface potential for each individual layer, and were also used to plot the surface potential differences of 2LG and 3LG relative to 1LG shown in Figure 3.

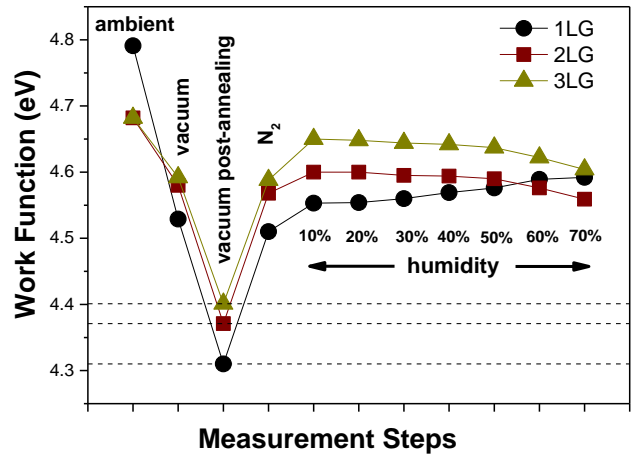

**Figure 4: Work function variation for 1, 2 and 3LG with the change in environment.**

The work function calculations for 1, 2 and 3LG assumed a constant work function for the tip, as no changes in free amplitude and fundamental oscillating frequency of the tip were detected with the change in humidity. Previous SKPM studies on graphite also found that the surface potential contribution of water adsorbed on the tip is significantly smaller than that adsorbed on the sample surface [29].

It can be seen that the work function values for 1, 2 and 3LG decrease with respect to ambient when the sample is annealed under vacuum, as expected with removal of p-dopants. A greater change in work function is observed for 1LG than for 2LG and 3LG on the transition from ambient to vacuum, indicating that 1LG is more sensitive to the change of environments than 2 and 3LG. When exposing the sample to increasing humidity levels, the work function of 1LG increases with humidity (10 - 70% RH). Interestingly, the work functions of 2LG and 3LG reveal a reverse trend compared to 1LG, showing a slow decrease in work function as the humidity increases. The work function values coincide for 1LG and 2LG at ~50% RH and for 1LG and 3LG at about ~70% RH (Fig. 4), indicating that the Fermi level is changing at a different rate for 1, 2 and 3LG with water adsorption. In summary, the SKPM results show that different layer numbers result in quantitative differences in water adsorption and change in environment. Furthermore, these differences are linked to changes in the physical and chemical properties of graphene's surface.

**Electrical transport measurements:** Surface potential measurements have been corroborated with electrical transport data recorded on graphene Hall bar devices containing 1LG and 2LG crosses of 760-nm width. The morphology and thickness of the device was determined using SKPM (details concerning device fabrication and transport measurement are presented in Methods). Transport measurements have been performed following a similar sequence, from ambient to vacuum, nitrogen and increasing RH in $N_2$, as for the SKPM experiments summarised in Figure 2. As shown in Figure 5, the carrier density (n-type) for both 1LG and 2LG increases in vacuum with respect to the values observed in ambient.

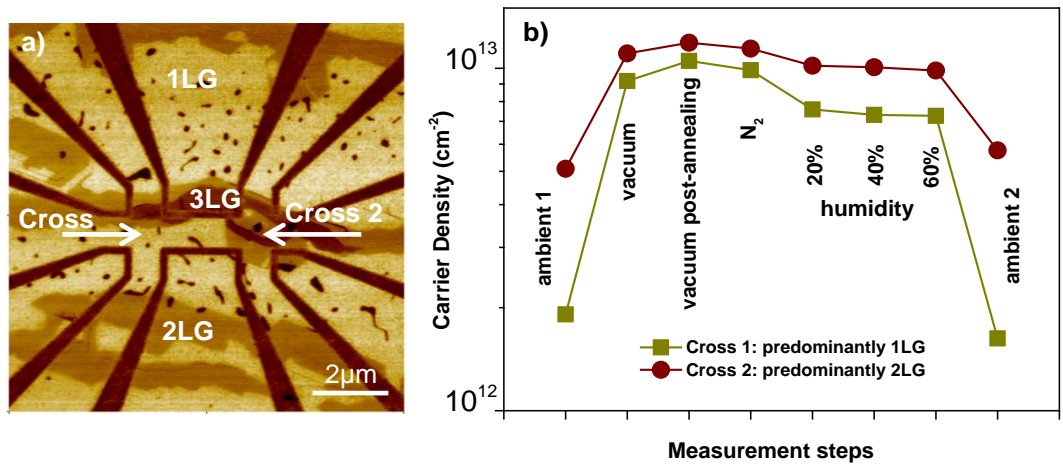

**Figure 5: (a) Surface potential image of the device used for transport measurements taken in vacuum, after temperature annealing at 150 °C. (b) Carrier density (n-type) extracted from transport measurements performed with the sample going through a similar sequence as for SKPM experiments in Figure 2.**

The change in carrier density is more pronounced for 1LG compared to 2LG, a trend that is followed throughout the entire sequence of measurements, although the largest carrier change occurs with the ambient-vacuum transition as also observed for the work function trend. A higher carrier density is determined for 2LG compared to 1LG for the entire measurement cycle. The environmental change in the carrier density follows the same trend as for the absolute values of surface potential for 1 and 2LG (shown in Supplementary Information section) and is generally in agreement with the recent work by Yang et al. treating doping by molecular adsorbates in ambient for epitaxial graphene [40]. Figure 5b illustrates that values of the carrier density at RH = 20 - 60% are significantly different from the values during ambient exposure for both 1 and 2LG.

Although the transport measurements were performed at a relatively high DC bias current ($I_{bias}$ = 50 µA), we observed no real change in the transport properties of the device when studied in the range 10-100 µA[41]. While there is no measurable increase of the temperature detected at $I_{bias}$ = 50 µA, a small temperature increase of 0.2° was measured in the vicinity of the device at $I_{bias}$ = 100 µA. Both these results together with a good thermal conductance of SiC confirm that there are no real inductive heating effects for micron-wide epitaxial graphene devices on 6$H$-SiC(0001).

Consistent with the SKPM results, the transport measurements show that the water adsorption behaviour is dependent on the layer number and that 1LG is more affected than 2LG upon exposure to RH, pointing to a different impact that surface absorbed water has on $E_F$ for 1LG relative to 2LG.

The transport data and the work function values determined by SKPM are used to schematically illustrate the energy band diagrams of 1LG and 2LG in Figure 6, where only the transitions from ambient to vacuum and to highest humidity level (RH = 70% in the transport measurements) are presented for simplicity.

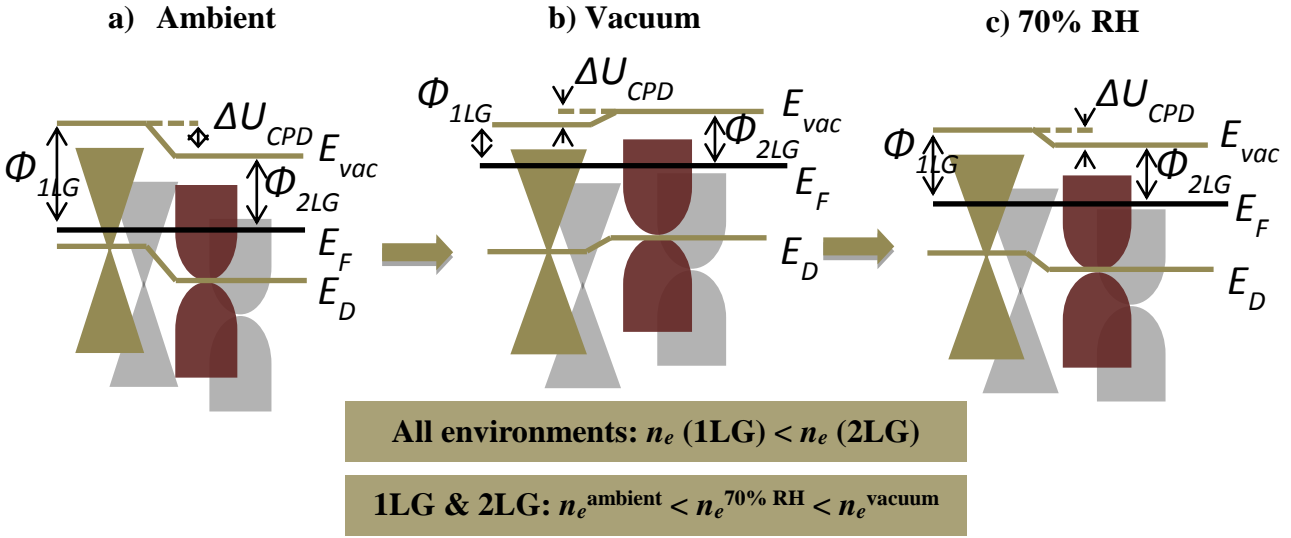

**Figure 6:** Schematic energy band diagram for 1LG and 2LG, illustrating respective changes in work function upon exposure to (a) air, (b) vacuum and (c) high humidity (RH = 70%).

A higher carrier (electron) density for both 1LG and 2LG is observed in vacuum compared to their respective ambient values, consistent with the work function values determined by SKPM. As discussed previously, the observed increase of the electron concentration in both 1LG and 2LG on ambient to vacuum transition is due to the desorption of environmental p-dopants and the increase for 1LG is larger than that of 2LG. With the addition of water, the sample shows an increase in the work function for both 1LG and 2LG relative to vacuum, consistent with an increased p-doping concentration and lower resulting carrier density for both 1LG and 2LG compared to the values measured in vacuum (Figure 6c). The difference in work function for 1LG on vacuum-humidity transition is larger than for 2LG. However, the doping concentration obtained for high RH levels does not reach the concentration observed in ambient, indicating that other factors in the ambient, in addition to water, are p-dopants and affect the surface potential of graphene. This observation together with qualitative differences in carrier densities for 1 and 2LG in all studied environments (Figure 5b) indicate higher sensitivity of 1LG to water vapour absorption as compared to 2LG. It is important to point out that $U_{CPD}$ values of 1LG and 2LG reproducibly restored to the initial values whenever re-exposing the sample to ambient, i.e. after vacuum or humidity treatment, as indicated in Figure 3e and also supported by the transport data in Figure 5b. These observations demonstrate that the properties of graphene are strongly influenced by external doping imposed by atmospheric adsorbates. The same trend exhibited by 2LG with the change in environment and exposure to humidity is also observed for 3LG.

In an attempt to clarify whether oxygen from ambient air has a greater effect on graphene than humidity, the sample was subjected to dry synthetic air containing 20% oxygen and 80% nitrogen, following pumping and annealing at 150°C under vacuum. No apparent effect of oxygen on the surface potential of 1LG, 2LG or 3LG was noted and the contact potential difference $U_{CPD}$ decreased only slightly in absolute value by ~3 mV compared to its vacuum value. The combined effect of oxygen and water was noted by previous works to be responsible for the hole doping of graphene [24, 27, 39]. We further tested the effect of humidity on the surface potential and transport data using synthetic air as a background gas (not shown). However, we observed the same trend as for humidity using $N_2$ gas (Fig. 3, 4 and 5), still not accounting for the full difference we observe relative to the starting ambient point.

**Adhesion mapping:** Since our study indicates a greater sensitivity of water for 1LG than 2LG, we further examine how the surface potential data correlates with the wetting behaviour of graphene domains of different thicknesses. For this purpose, adhesion mapping has been carried out by recording force-distance curves at every pixel on a preselected region on the sample. The experiments have been performed with the sample immersed in de-ionised (DI) water (resistivity 18.2 MΩ·cm) in order to avoid the effect of the water contamination layer usually present on surfaces scanned in ambient, giving rise to a liquid capillary between the probe and the sample [30] that is known to cause relatively large attractive interactions. In the present work, OTS functionalized tips have been used to map the spatial variation of attractive forces between hydrophobic tips and graphene in DI water. The resulting adhesion map, illustrated in Figure 7, clearly evidences the presence of 1LG and 2LG domains, consistent with the surface potential measurements presented in Figure 2.

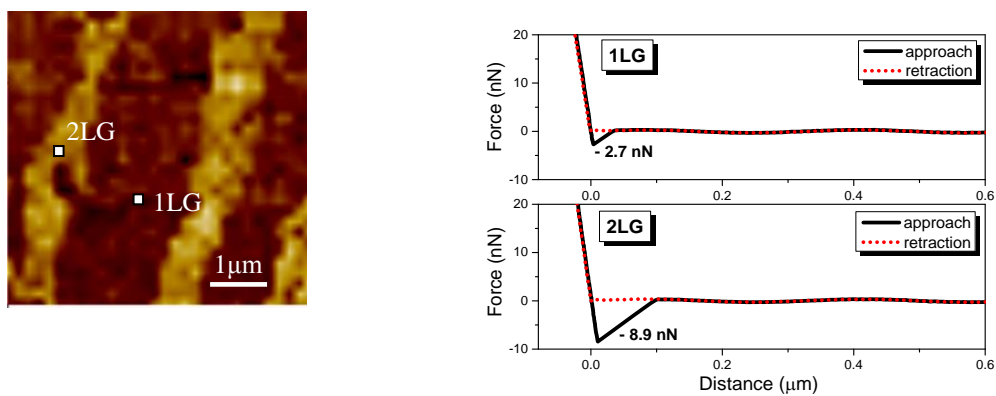

**Figure 7: a) Adhesion map obtained on the sample immersed in DI water, using OTS functionalised probe. b) Associated force-distance curves extracted from selected locations on 1LG and 2LG domains.**

Test samples with a pattern of hydrophobic OTS areas and hydrophilic gold areas confirmed the larger adhesion force on hydrophobic areas (see Supplementary Information) when using OTS functionalized tips. This indicates that a larger adhesion force is measured on more hydrophobic areas of the sample (i.e. 2LG), due to the hydrophobic interaction, compared to less hydrophobic areas (i.e. 1LG). As highlighted in the force-distance curves in Figure 7b, a larger adhesion force is consistently measured on 2LG as compared to 1LG domains, suggesting that 1LG is less hydrophobic. Attractive forces of the order of $(7.6 \pm 1.6)$ nN have been measured for 2LG domains and $(4.1 \pm 1.4)$ nN for 1LG. In humid air, the lower hydrophobicity is associated with a greater capillary force on 1LG compared to 2LG and is consistent with current SKPM and electrical transport observations of enhanced water sensitivity for 1LG relative to 2LG.

**Discussion**: Overall, both SKPM and transport measurements show that the thickness of graphene results in quantitative differences in water adsorption, that are linked to changes in graphene's surface physical and chemical properties. These properties change more for 1LG than for 2 and 3LG and water adsorption is greater on 1LG than on thicker graphene.

The lack of sensitivity of 2 and 3LG relative to 1LG is most likely due to the different electronic structure of single and multilayer graphene as shown by previous studies demonstrating that the electronic band dispersion near the Fermi level, and consequently the nature of the charge carriers, is highly sensitive to the number of layers, the stacking geometry and the interlayer interaction [35, 36, 37]. The $\pi$ bands of a single graphene layer do not exhibit dispersion with the out-of-plane electron momentum. However, each added layer affects the topology of the $\pi$ bands, and an additional band occurs due to the interaction between layers, as shown by angle-resolved photoemission spectroscopy (ARPES) [36]. From ARPES measurements of epitaxial graphene on SiC, Ohta et al. extracted tight binding parameters, including interlayer hopping integrals and $E_F$ =440, 300, 210, and 150 meV for 1, 2, 3 and 4 LG, respectively [36]. Since the total sheet charge density did not change much, it was concluded that interlayer screening results in a charge redistribution, with an effective screening length of 1.4 and 1.9 Å for 3 and 4 LG, respectively, that increases further for thicker layers and graphite. Furthermore, DFT calculations showed that water molecules are less bound to 2LG than to 1LG [16], indicating a binding energy of 1.30 eV per water molecule for 2LG, compared to 1.94 eV for 1LG.

Both SKPM and transport measurements show that neither water nor oxygen alone has a dominating effect on the electronic properties of graphene. The same can be inferred about their combined contribution. But, since the case of the ambient exposure is so different from the controlled exposures using high purity gases, the implication is that some other (unknown) impurity (on their own or in combination with water and oxygen) is responsible for the difference that is measured in ambient relative to high humidity environment. Several previous theoretical studies showed that adsorbed water has very little effect on the electronic structure of graphene and that physisorption is generally observed at the water-graphene interface [7, 17, 19]. It was shown by DFT calculations [17] that for small water clusters, charge transfer between graphene and water is very small and does not influence the density of charge carriers due to the fact that water molecules orient their dipole moments in opposite directions, so that they cancel on average. The discrepancy with experimental studies showing doping effects due to water adsorbates is likely to originate from the effect of the supporting substrate, as most of these studies use exfoliated graphene on $SiO_2$ substrates. The substrate surface quality and chemical cleanliness were shown to play a crucial role for observation of doping by water adsorbates on graphene [19]. This indicates that more complex mechanisms might be involved than the simple interaction of graphene with single water molecules, for example, through the water-substrate interaction, as shown by Wehling et al. [19]. The dipole moments of water adsorbates can shift the substrate impurity bands and change their hybridisation with the graphene bands, leading to doping of graphene on thermally oxidised silicon substrates [19]. It was also postulated that oxygen and water adsorbates compensate for negatively-charged impurities in the supporting $SiO_2$/Si substrate: for graphene sheets consisting mainly of p-type regions (corresponding to negatively-charged impurities in the substrate), the adsorbate layer tends to screen the substrate impurities and assists in mitigation of carrier scattering [9]. However, it is not clear whether the SiC substrate or the buffer layer at the graphene/SiC interface, controlling the electrostatic conditions of epitaxial graphene, play a similar role in mediating the interaction of water with epitaxial graphene.

Apart from water and oxygen responsible for the hole doping of graphene, the difference in the carrier density and work function values of graphene in ambient relative to high humidity suggests that other p-type doping contaminants commonly present in air have an additional effect on graphene. Likely candidates could be volatile hydrocarbon compounds that have been shown to significantly affect the wettability of graphitic surfaces through a large change in the water contact-angle with graphene, i.e. from 37° for a clean surface to about 80° for a hydrocarbon-contaminated one [31].

**Conclusions:** In summary, the electronic properties of graphene in controlled humidity environment have been examined in this study. The study shows that graphene's physical and chemical properties are sensitive to water vapour. Moreover, graphene's response to water and change in environment is strongly thickness dependent, with different layer numbers resulting in quantitative differences in water adsorption. The differences in water adsorption are linked to changes in graphene's physical and chemical properties. The results indicate increased water sensitivity with decreasing graphene thickness, with 1LG being the most affected by water and the change in environment. This observation is consistent with adhesion mapping measurements indicating that 1LG is less hydrophobic than 2LG. The study shows that work function and carrier density of 1, 2 and 3LG restore to the initial ambient values after the sequence of exposures involving vacuum and RH. The change in work function and carrier density with the change in environment is more pronounced for 1LG than for 2LG, with 2LG consistently showing higher carrier density than 1LG for all examined environments. The RH exposure demonstrates a doping effect (p-type) that partly compensates the intrinsic n-type nature of the sample, as the overall carrier density decreases with increasing RH exposure. The work function of 1LG increases with increasing humidity level and the change in work function and doping is different in 1LG and 2LG for the same RH exposure.

Unexpectantly, our study finds that other contaminants commonly present in ambient air (such as volatile organic compounds or trace impurity gases) have a significant effect on the work function and doping of graphene. This is because the observed changes in work function and doping far exceed the changes found using pure Nitrogen and Oxygen gases, or via introduced humidity created from pure water vapour. This observation implies that real sensors need to be made immune to the effects of humidity and other constituents in the ambient, as selectivity is an essential ingredient of graphene sensors designed to work in in the ambient. Hence, additional experiments to understand the response to environmental effects are required so that realistic models can be formulated to insure the desired sensitivity of the sensor to the target molecule is obtained. Our work will provide a foundation for developing models for realistic sensors which could be extended to sensor designs that deliberately minimize the effects of the ambient.

**Methods**

1. The substrate (II-VI, Inc.) was ca. 8x8 mm$^2$ of semiinsulating (0001)6*H*-SiC (resistivity >10$^{10}$ ohm cm) misoriented ~ 0.05° from the basal plane mainly in the (11-20) direction. Graphene was synthesized via Si sublimation from SiC using an overpressure of an inert gas. The substrates were etched in H$_2$ at 200 mbar using a ramp from room temperature to 1580°C to remove polishing damage. At the end of the ramp, the H$_2$ was evacuated and Ar added to a pressure of 100 mbar (the transition takes about 2 minutes). The graphene was then synthesized at 1580°C for 25 min in the Ar. Afterwards, the sample was cooled in Ar to 800°C [38]. Two or more layers of graphene formed in this fashion are known to be Bernal stacked [36].

2. **Raman spectroscopy:** Raman intensity maps were obtained using a Horiba Jobin-Yvon HR800 System. A 532-nm excitation was focused onto the sample through a 100x objective with 1 mW power incident and data were taken with a spectral resolution of (3.1 ± 0.4) cm$^{-1}$ and XY resolution of (0.4 ± 0.1) μm. The raw data were normalised with respect to the maximum of the two-phonon mode of SiC at ~ 1514 cm$^{-1}$.

3. **Scanning Kelvin probe microscopy** (SKPM) experiments in ambient, vacuum, N$_2$ and humidity exposure were conducted on an NT-MDT NTEGRA Aura SPM system, using Bruker highly doped Si probes (PFQNE-AL) with a force constant ~ 0.9 N/m and resonant frequency of f$_0$ ~ 300 kHz. Frequency-modulated SKPM (FM-SKPM) technique operated in a single pass mode has been used in all measurements. FM-SKPM operates by detecting the force gradient (dF/dz), which results in changes to the resonance frequency of the cantilever. In this technique, an AC voltage with a lower frequency ($f_{mod}$=3 kHz) than that of the resonant frequency of the cantilever is applied to the probe, inducing a frequency shift. The feedback loop of FM-KPFM monitors the side modes, $f_0 \pm f_{mod}$, and compensates the mode frequency by applying an offset DC voltage which is recorded to obtain a surface potential map. Since FM-KPFM detects the force gradient using the frequency shift, it can achieve spatial resolution of <20 nm, which is limited only by the tip diameter, enabling the determination of the number of graphene layers of the sample with great accuracy.

For 1, 2 and 3LG work function calculations, the work function of the SKPM tip was determined using calibration against a gold reference sample. The work function of the gold sample was measured using ultra-violet photoelectron spectroscopy [32]. The value obtained for the work function of the tip is (4.100 ± 0.003eV). It is important to note that a different value for the work function of the tip, although changing the absolute value of work functions we report for 1, 2 and 3LG, would nevertheless result in the same trend that we observe upon the change in environment. The errors quoted in the manuscript for the work functions of 1, 2 and 3LG have been determined from the standard deviation of the surface potential line profiles.

**SKPM under different environments**: FM-SKPM measurements in ambient air and vacuum (P ~1×10$^{-5}$ mbar) were performed as described above. Measurements under vacuum were performed after the sample cooled down to room temperature following the sample annealing at 150°C. Controlling the SPM chamber environment to $N_2$ and synthetic air simply required regulating the gas flow using an inlet valve. For humidity exposure, RH was increased in a step-wise manner from 10% to 70% and the surface potential measurements were carried out immediately after the chamber reached the target humidity level and the time required for equilibrium to be established at low RH: ~5 mins, and up to ~ 30 mins for 70% RH. The measurement time to acquire one complete scan is 15 min. For each RH step, subsequent measurements were taken up to 1h following each exposure and showed no change in surface potential, suggesting that the humidity effect was relatively quickly saturated on the surface of the graphene (less than 5 min. for 10% RH exposure).

**4. Adhesion mapping:** Employing a Cypher AFM system (Asylum Research, CA), the adhesion forces between chemically functionalized AFM probes and graphene surfaces were measured. The AFM system was fitted with a superluminescent diode to minimize signal oscillations resulting from optical interference of a non-zero fraction of light reflected off the sample surface with the light reflected off the cantilever. The AFM cantilevers used were of type CSC38 with aluminium reflective coating (Mikromasch Europe, Germany). Of the three rectangular silicon cantilevers of different lengths and spring constants available on a CSC38 chip, either cantilevers of ~300 or ~350 μm length were used. Typical spring constant values specified by the manufacturer are ~50 and ~30 pN/nm, respectively. Measured values of the spring constant were in the range of ~80 to 200 pN/nm and ~80 to 150 pN/nm, respectively. To allow hydrophobic behavior of the AFM tip, the cantilevers were functionalized with octadecyltrichlorosilane (OTS). Prior to immersing into a solution of OTS in toluene, the cantilevers were treated in a UV/ozone cleaner for ~20 min. Immediately after the immersion, the cantilevers were washed three times, in chloroform, acetone and again in chloroform. The concentration of the OTS-toluene solution was ~$10^{-3}$ mol/L. The AFM force measurements were undertaken in de-ionized (DI) water, by immersing the cantilever into a drop of ~80 μL volume on top of the graphene sample. The force-displacement curves were run at a scan rate of ~0.81 Hz and a z-velocity of ~1.93 μm/s. They were recorded over arrays of points and the magnitude of the pull-off peaks occurring upon retraction of the probe was analyzed. Typically, the pixel number of the resulting force maps was 64x64. To check the functionalization of the cantilevers and to characterize related adhesion contrasts, test samples with an array of gold squares on top of a silicon wafer surface were prepared.

**5. Device fabrication**: Graphene devices in Hall bar geometry were fabricated from epitaxial graphene on Si-face 6*H*-SiC with electron beam lithography using Poly(methyl methacrylate) and ZEP520 (ZEON Chemicals) positive resists (thickness: 250 and 200 nm, respectively), oxygen plasma etching (one minute) and electron beam physical vapour deposition of Cr/Au (5/100 nm). Graphene samples used for device fabrication were from the same batch of the graphene films used for the SKPM described in the paper. Further details on the fabrication process are reported in Ref. [33]. Using these fabrication processes, a 760 nm wide Hall bar device with cross 1 covered by 1LG (~70%) and 2LG (~30%), cross 2 covered by 2LG (~70%) and 3LG (~30%) and channel of 2.6 µm length covered by 1LG (~56%), 2LG (~30%) and 3LG (14%) was obtained (see Figure 5 for SKPM map of the device in the main text). Residues due to contamination from resists and solvents as a result of the fabrication process have been removed by mechanically scraping the residues from side-to-side using soft cantilevers in contact-mode AFM to avoid damaging the graphene device [32, 33]. Figure 5a in the manuscript shows the SKPM map of the device following the cleaning process.

**6. Transport measurements:** The transport properties of the device were investigated using the AC Hall effect and 4-point resistance measurements. The AC Hall effect was investigated by applying an out-of-plane AC magnetic field ($B_{AC}$ = 5 mT at 126 Hz) [34]. The resulting transverse AC signal of the DC current biased ($I_{bias}$ = 50 µA) Hall bar device contains contributions from the AC Hall voltage (in-phase) and inductance effects (90° out-of-phase). Using Stanford Research Systems SR830 lock-in amplifiers and the $I_{bias}$ reversal technique, the AC Hall voltage ($V_H$) was accurately separated by measuring only the in-phase component of the AC signal at +/− $I_{bias}$. The resistance of the graphene channel ($R_4$) was determined using the 4-point technique, i.e., by applying +/− $I_{bias}$ and measuring the voltage drop from cross 1 to cross 2. The current reversal during the $R_4$ measurement eliminates the thermal electric voltage offset. From these measurements, the carrier density $\left(n = \frac{I_{bias} B_{AC}}{eV_H}\right)$ and carrier mobility $\left(\mu = \frac{L}{W}\frac{1}{enR_4}\right)$ were obtained, where *e* is the electronic charge and L/W is the device length/width ratio (~3.52).

ASSOCIATED CONTENT: Surface potential values for 1, 2 and 3LG with the change in environment and adhesion measurements of the test sample are included in the Supplementary Information section.

**Supplementary Information**

Surface potential measurements for one-, two- and three-layer graphene (1, 2, and 3LG) show the same trend described by the carrier concentration with the change in environment (Figure 5b in the manuscript). The surface potential values given in Table S1 are plotted in Figure S1:

**Table S1: Surface potential values of 1, 2, and 3LG for each environment extracted from images presented in Figure 2 in the manuscript.**

| Environment | SP $_{1LG}$ / eV (±0.020eV) | SP $_{2LG}$ / eV (±0.020eV) | SP $_{3LG}$ / eV (±0.020eV) |
|---|---|---|---|
| Ambient | -0.691 | -0.582 | -0.582 |
| Vacuum | -0.429 | -0.48 | -0.492 |
| Vacuum post-annealing | -0.21 | -0.271 | -0.301 |
| N$_2$ | -0.41 | -0.468 | -0.488 |
| 10% RH | -0.453 | -0.5 | -0.55 |
| 20% RH | -0.454 | -0.5 | -0.548 |
| 30% RH | -0.46 | -0.495 | -0.544 |
| 40% RH | -0.469 | -0.494 | -0.542 |
| 50% RH | -0.476 | -0.49 | -0.537 |
| 60% RH | -0.489 | -0.476 | -0.522 |
| 70% RH | -0.492 | -0.459 | -0.504 |

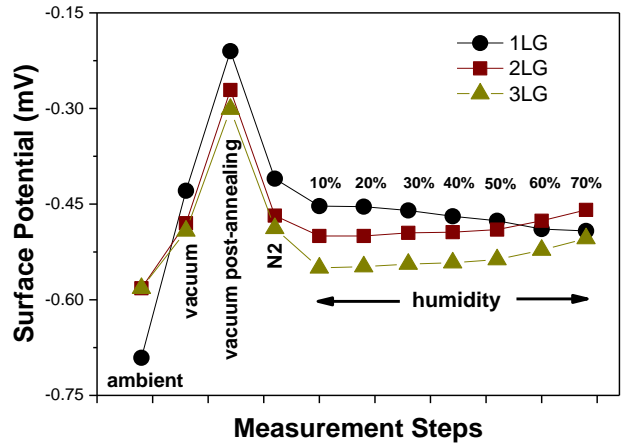

**Figure S1:** Surface potential for 1, 2 and 3LG with the change in environment, from ambient to vacuum, and to increasing humidity levels.

**Table S2:** Work function values derived from SKPM measurements on 1, 2 and 3LG exposed to ambient, vacuum, low and high humidity levels:

| Environment | $\Phi_{1LG}$ (eV) | $\Phi_{2LG}$ (eV) | $\Phi_{3LG}$ (eV) |
|---|---|---|---|
| Ambient | 4.79 ± 0.02 | 4.68 ± 0.02 | 4.69 ± 0.02 |
| Vacuum post-annealing | 4.31 ± 0.02 | 4.37 ± 0.02 | 4.40 ± 0.02 |
| 10% RH | 4.55 ± 0.02 | 4.60 ± 0.02 | 4.65 ± 0.02 |
| 70% RH | 4.59 ± 0.03 | 4.56 ± 0.03 | 4.6 ± 0.03 |

**Adhesion measurements of test samples:**

Adhesion force mapping on test samples consisting of a pattern of hydrophobic octadecyltrichlorosilane (OTS) areas and hydrophilic gold areas, shown in Figure S2(a), confirmed the larger adhesion force on hydrophobic areas (Figure S2b). The histogram associated with the adhesion map shows the range of pull-off forces associated with OTS and Au areas (Figure S2c).

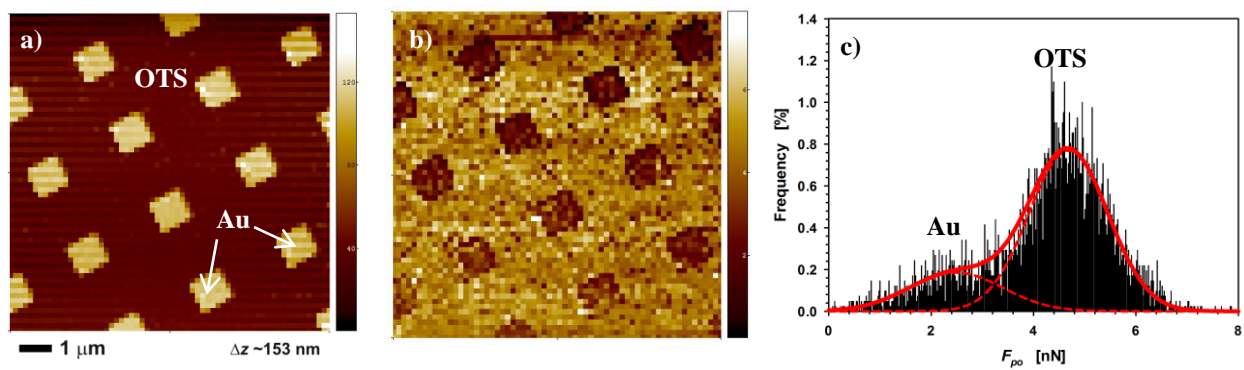

**Figure S2: (a) Height map of OTS and Au pattern, (b) Corresponding adhesion map, (c) Histogram of the adhesion map with the Gaussian fits to the peaks indicated by red lines.**

ACKNOWLEDGEMENT: The work was supported by the NMS under IRD Graphene Project, Graphene Flagship (No. CNECT-ICT-604391). Work at the Naval Research Laboratory was supported by the Office of Naval Research.